\documentclass[12pt]{article}
\usepackage{amssymb,amsfonts,epsfig}

\newcommand{\beq}{\begin{equation}}
\newcommand{\beqn}{\begin{eqnarray}}
\newcommand{\eeq}{\end{equation}}
\newcommand{\eeqn}{\end{eqnarray}}

\def\ntwo{${\mathcal N}=2\,$}

\def\none{${\mathcal N}=1\,$}

\begin{document}

%%%%%%%%%%%%%%%%%%%%%%%%%%%%%%%%%% definitions

% \def\longvert{{\rule[-2mm]{0.1mm}{7mm}}\,}

\newcommand{\nc}{\newcommand}
\def \foot {\footnote}
\def\stroke{\vrule height8pt width0.4pt depth-0.1pt}
\def\topfleck{\vrule height8pt width0.5pt depth-5.9pt}
\def\botfleck{\vrule height2pt width0.5pt depth0.1pt}
\def\Zmath{\vcenter{\hbox{\numbers\rlap{\rlap{Z}\kern
0.8pt\topfleck}\kern 2.2pt\rlap Z\kern 6pt\botfleck\kern 1pt}}}
\def\Qmath{\vcenter{\hbox{\upright\rlap{\rlap{Q}\kern
3.8pt\stroke}\phantom{Q}}}}
\def\Nmath{\vcenter{\hbox{\upright\rlap{I}\kern 1.7pt N}}}
\def\Cmath{\vcenter{\hbox{\upright\rlap{\rlap{C}\kern
3.8pt\stroke}\phantom{C}}}}
\def\Rmath{\vcenter{\hbox{\upright\rlap{I}\kern 1.7pt R}}}
\def\Z{\ifmmode\Zmath\else$\Zmath$\fi}
\def\Q{\ifmmode\Qmath\else$\Qmath$\fi}
\def\N{\ifmmode\Nmath\else$\Nmath$\fi}
\def\C{\ifmmode\Cmath\else$\Cmath$\fi}
\def\R{\ifmmode\Rmath\else$\Rmath$\fi}

%%%%%%%%%%%%%%%%%%%%%%%%%%%%%%%%%%%%%%%%%%%%%%%%%%%%%%%%%%%%%%%%%%%%

\def\baselinestretch{1.0}

\begin{titlepage}

\begin{flushright}
FTPI-MINN-08/30, UMN-TH-2711/08\\
August 13, 2008
\end{flushright}

\vspace{0.5cm}

\begin{center}

{\Large \bf Revisiting Critical Vortices \\[2mm]
in Three-Dimensional SQED }

\end{center}

\vspace{1.5cm}

\begin{center}
{{\bf S. \"Olmez}$^{\,a}$, {\bf M.~Shifman}$^{\,b\,,c}$ }

\vspace{2mm}

$^a${\it Department of Physics and Astronomy, University of
Minnesota, Minneapolis, MN 55455, USA}\\[2mm]
$^b${\it  William I. Fine Theoretical Physics Institute, University
of Minnesota, Minneapolis, MN 55455, USA}\\[2mm]
$^c$ {\it Laboratoire de Physique Th\'eorique\footnote{Unit\'e Mixte
de Recherche du CNRS,  (UMR 8627).}
Universit\'e de Paris-Sud XI\\
B\^atiment 210, F-91405 Orsay C\'edex, FRANCE}

\end {center}

\begin{center}
{\bf Abstract}
\end{center}

We consider renormalization of the central charge and the mass of
the ${\cal N}=2$ supersymmetric Abelian vortices in $2+1$
dimensions. We obtain ${\cal N}=2$ supersymmetric theory in $2+1$
dimensions by dimensionally reducing the ${\cal N}=1$ SQED in $3+1$
dimensions with two chiral fields carrying opposite charges.
Then we introduce a mass for one of the
matter multiplets
without breaking ${\cal N}=2$ supersymmetry.  This  massive multiplet  is viewed as a regulator in the large mass limit. We show that the mass and the central charge of the vortex get the same nonvanishing quantum corrections, which preserves BPS saturation at the quantum level.
Comparison with the operator form of the central extension
exhibits fractionalization of a global U(1) charge; it becomes $\pm 1/2$
for the minimal vortex. The very fact of the mass and charge
renormalization is due to a ``reflection" of an unbalanced number of
the fermion and boson zero modes on the vortex
in the regulator sector.\\
\mbox{}
\\
\mbox{}

\end{titlepage}

\section{Introduction}
\label{intro}

${\cal N}=2$ supersymmetric QED with the Fayet--Iliopoulos term in
$2+1$ dimensions supports Abrikosov--Nielsen--Olesen (ANO) vortices
\cite{Schmidt,Edelstein}. These classical solutions are 1/2-BPS
saturated (two out of four supercharges are conserved). Quantum
corrections to the vortex mass and central charge were discussed in
the literature more than once. It is firmly established
\cite{Rebhan1} that there are two fermion zero modes on the vortex
implying that the supermultiplet to which the vortex belongs is
two-dimensional. This is a short supermultiplet. Hence, the
classical BPS saturation cannot be lost in loops.

Particular implementation of the vortex BPS saturation turned out to
be a contentious issue, almost to the same extent as it had happened
with two-dimensional kinks in \none models (for reviews see
\cite{SVV}, Sect. 3.1 in \cite{shifmanyung},  and \cite{RebhanPN}).
The authors of \cite{Schmidt} and
\cite{B.H.Lee and H. Min} obtained a vanishing quantum correction to
the vortex mass using the following eigenvalue densities:
\begin{equation}
\label{index}
   n_B (w)-n_F(w)\propto\delta(w)\, ,
\end{equation}
where $n_{B(F)}$ is the bosonic (fermionic) density of states. The
vanishing mass correction ensues since
\begin{equation}
\label{index}
   \Delta M_v  \propto \int d w \,\left(n_B (w)-n_F(w)\right)\,w=0\,.
\end{equation}
Since the vortex mass $M_v$ is proportional to the Fayet--Iliopoulos
(FI) parameter $\xi$, and $\xi$ {\em is} renormalized in one loop,
the above result caused a problem.

Later new calculations of the vortex mass were undertaken and a
nonvanishing one-loop correction to the vortex mass was reported in
\cite{vassil,Shizuya}. It was shown \cite{Rebhan1} that the central
charge also gets a correction, so that the BPS saturation of the
vortex persists at the one-loop level. However, the (dimensional)
regularization that was used in the most detailed paper
\cite{Rebhan1}, expressly written to discuss three-dimensional
supersymmetric vortices, does not allow one to treat in a
straightforward manner the Chern--Simons (CS) term, whose role in
the problem at hand is important. In this paper we use another
regularization method in which the CS term naturally appears in the
limit of large regulator mass. This mass is also crucial in the
operator form of the centrally extended algebra which we derive at
one loop. Our operator expression for the central extension includes
the Noether charge (\ref{Momentum along 3rd axis}).

 Here we would like to
close these gaps. In this paper we revisit the issue using a
physically motivated regularization which is absolutely transparent.
We recalculate the renormalization of the vortex mass at one loop
\beq M_{v,\rm R} = 2\pi\,\left(\xi_{\rm R} - \frac{m}{4\pi} \right)
\label{1} \eeq
 and the
one-loop effect in the central charge. (Here $\xi_{\rm R}$ is the
renormalized value of the FI parameter, $m$ is the matter field
mass, $$m =e\sqrt{2\xi_R}\,,$$ and the subscript R stands for
renormalized.) The above result is in agreement with the previous
calculations \cite{Rebhan1,vassil}.
 Needless to say, our direct calculation confirms BPS saturation,
$M_{v,\rm R} =|Z_{\rm R}|$. Moreover, it demonstrates that, in the
limit of the large regulator mass, regulator's role is taken over by the
Chern--Simons term. A new finding obtained by comparing
the central charge calculation with the operator form of the
central extension is a U(1) global charge fractionalization.
The operator expression for the central extension which we derived in our regularization
is presented in Eqs.~(\ref{susy algebra}) and (\ref{Momentum along 3rd axis}).
Then we discuss the central charge/vortex mass
renormalization to all orders in perturbation theory, see Eq.~(\ref{54ors}).

\ntwo SQED Lagrangian in $2+1$ dimensions (four supercharges) can be
obtained by dimensional reduction of ${\mathcal N}=1$ supersymmetric
Lagrangian in $3+1$ dimensions. In order to have a well defined
anomaly-free  SQED in four dimensions, one has to have two matter
superfields, say $\Phi$ and $\tilde{\Phi}$, with the opposite
charges. Since there is no chirality in three dimensions, in
three-dimensional SQED, in principle, it is sufficient to keep a
single superfield (say, $\Phi$), while $\tilde{\Phi}$ can be
eliminated. This is a minimal setup which is routinely considered.
The four-dimensional anomaly is reflected in three dimensions in the
form of a ``parity anomaly'' \cite{semenoff, redlich} and the
emergence of the Chern--Simons term, as will be explained
momentarily.

When we speak of eliminating $\tilde{\Phi}$ we should be careful.
Eliminating does not mean discarding. As was briefly discussed in
\cite{shifmanyung} (Sect. 3.2), a perfectly safe method of getting
rid of $\tilde{\Phi}$ is to make the tilded fields heavy. Then the
corresponding supermultiplet decouples and does not appear in the
low-energy theory. It leaves a trace, however, in the form of the
Chern--Simons term \cite{semenoff,redlich}, as shown in Sect. 4.

There is a well-known method of making the tilded fields heavy
without altering the masses of the untilded fields. It works in
three dimensions. One can introduce a ``real" mass  $\tilde m$
\cite{AHISS} (a
three-dimensional analog of the twisted mass in two dimensions
\cite{twisted}) without breaking \ntwo supersymmetry of
three-dimensional SQED. The real mass corresponds to a constant
background vector field along the reduced direction.

When the masses of the tilded and
untilded fields are equal, the renormalization of the FI term
vanishes \cite{noxi}, and so do quantum corrections to the vortex
mass. When we make the tilded fields heavy, $\tilde{m}\gg
e\sqrt\xi$,  effectively they become physical regulators. As long as
we keep their mass $\tilde{m}$ large but finite it acts as an
ultraviolet cut-off in loop integrals. All one-loop corrections,
including the linearly divergent part, become well-defined and
perfectly transparent. We have a smooth transition as we eventually send
$\tilde m$ to infinity.

Our analysis is organized as follows. In Sect.~\ref{descr}
we describe our basic model obtained from four-dimensional SQED by
reducing one of the spatial dimensions. We introduce
the real mass $\tilde m$, to be treated as a free parameter,
for the ``second" chiral superfield. Section \ref{section Quantum Corrections},
carrying the main
weight of this work, is devoted to quantum corrections to the
central charge and vortex mass. The operator form of the central extension is discussed in
detail in this section.
In Sec.~\ref{q} we consider a global charge fractionalization and a
related question of Chern--Simons.

\section{Description of the model and classical results}
\label{descr}

Our starting point is ${\mathcal N}=1$ SQED in $3+1$ dimensions with
two chiral matter superfields $\Phi$ and $\tilde\Phi$ and the
Fayet--Iliopoulos term. It has four conserved supercharges. The
corresponding Lagrangian is
\begin{eqnarray}
\label{lagrangian n=1} {\mathcal L } &=& \left\{ \frac{1}{4
e^2}\int\!{\rm d}^2\theta \, W_\alpha W^\alpha  + {\rm H.c.}\right\}
+ \int \!{\rm d}^4\theta \,\Phi^*\, e^{V}\, \Phi
\nonumber\\[3mm]
&&+ \int \!{\rm d}^4\theta \,\tilde{\Phi}^*\, e^{-V}\, \tilde{\Phi}
- \, \xi  \int\! {\rm d}^2\theta {\rm d}^2 \theta^\dagger \,V(\!
x,\theta , \theta^\dagger ) \, ,
\label{sqed}
\end{eqnarray}
where ${W}_{\alpha}$ is the gauge field multiplet,
\begin{equation}
{W}_{\alpha} = \frac{1}{8}\;\bar{D}^2\, D_{\alpha } V =
 \lambda_{\alpha} -\theta_{\alpha}D - i\theta^{\beta}\,
F_{\alpha\beta} +
i\theta^2{\partial}_{\alpha\dot\alpha}{\lambda}^{\dagger\,\dot\alpha}
\, .
\end{equation}
In order to get ${\mathcal N}=2$ supersymmetry in $2+1$ dimensions
we compactify one of the dimensions, say the third axis,  keeping
the zero Kaluza--Klein modes and discarding nonzero ones. To
introduce the tilded field mass we introduce a constant background
gauge field along the compactified axis, $V_{\mathrm{bg}}$, where
the subscript $\mathrm{bg}$ means background. In terms of the
components we have
\begin{equation}\label{xx}
   V_{\mathrm{bg}}= \theta^\dagger
    \gamma^0 \gamma^\mu\theta V^{\mathrm{bg}}_\mu,
\end{equation}
 $\gamma$-matrices are  defined in
Eq.~(\ref{gamma}) below.  The background vector field is chosen to
be a constant field along the compactified axis, i.e.
$V_{\mathrm{bg}}^\mu=2 \tilde m \,\delta^\mu_3$. It is important to
note that this is a new auxiliary field, rather than the expectation
value of the original photon field. This background is coupled to
$\tilde \Phi$ only, with the charge $-1$. Then the Lagrangian takes
the form
\begin{eqnarray}
\label{lagrangian n=1} {\mathcal L } &=& \left\{ \frac{1}{4
e^2}\int\!{\rm d}^2\theta \, W_\alpha W^\alpha  + {\rm H.c.}\right\}
+ \int \!{\rm d}^4\theta \,\Phi^*\, e^{V}\, \Phi
\nonumber\\[3mm]
&&+ \int \!{\rm d}^4\theta \,\tilde{\Phi}^*\, e^{-V-V_{\mathrm{bg}
}} \tilde{\Phi} - \, \xi  \int\! {\rm d}^2\theta {\rm d}^2
\theta^\dagger \,V(\! x,\theta , \theta^\dagger ),\label{sqed2}
\end{eqnarray}
Upon introduction of the constant background field, $\tilde \Phi$
multiplet becomes massive whereas $\Phi$ multiplet is not affected,
since it is chosen to be neutral with respect to the background
field. It is clear that the kinetic term for the gauge multiplet is
not affected,  and similarly, the Fayet-Iliopoulos term remains the
same since the superspace integral $\int d^4\theta V$ does not
vanish only for the last component of the superfield $V$.

After compactification of the third axis, we
get the following bosonic and fermionic Lagrangians in terms of
the component fields (in the Wess--Zumino gauge):
\begin{eqnarray}
\label{bosonic lagrangian}
{\cal L}_{B} &=& -\frac{1}{4 e^2}\, F_{\mu\nu}F^{\mu\nu}+ {\mathcal
D}^\mu\tilde {\phi}^*\, {\mathcal D}_\mu \tilde {\phi} + {\mathcal
D}^\mu\phi^*\, {\mathcal D}_\mu \phi+\frac{1}{2
e^2}\,\left(\partial_\mu\,N\right)^2
\nonumber\\[2mm]
&+& \frac{1}{2 e^2}\,D^2 -\, \xi\,D + \,D( \phi^*\phi- \tilde
{\phi}^*\tilde {\phi})-  N^2 \phi^* \,\phi -(\tilde m+N)^2
\,\tilde{{\phi}}^*\,\tilde{\phi}\,,
\nonumber\\[3mm]
 {\cal L}_{F}&=&\frac{1}{e^2}\bar\lambda
\,i\,\!{\not\!\partial}\,\lambda  + \bar\psi
\,i\,\!{\not\!\!{\mathcal D}}\,\psi +\bar{\tilde\psi
}\,i\,\!{\not\!\!{\mathcal D}}\,\tilde\psi +
  N\, \bar\psi\,\psi -(\tilde m+N)\, \bar{\tilde\psi}\,\tilde\psi
\nonumber\\[2mm]
&+& i \sqrt{2}\left[\left(\bar\lambda\,\psi\phi^*-\bar\psi \lambda\,
\phi\right)\right]-i\sqrt{2} \left[(\bar\lambda\,\tilde \psi )\tilde
{\phi}^*- (\bar{\tilde \psi}\lambda )\tilde {\phi}\right],
\label{N=2 2+1 dimensions}
\end{eqnarray}
where $N=-\,A_3$ is a real pseudoscalar field, and
$$
 i{\mathcal D}_\mu \phi=\left( i \partial_\mu +  A_\mu\right)\phi\,,\quad
 i{\mathcal D}_\mu \tilde\phi=\left( i \partial_\mu -  A_\mu\right)\tilde\phi\
\,.
 $$
Moreover, $D$ is an auxiliary field, which can be eliminated via its
equation of motion. The Lagrangian (\ref{N=2 2+1 dimensions}) is
invariant under the following supersymmetry transformations,
\begin{eqnarray}
\label{susy transformations}
  \delta \phi&=&\sqrt 2 \bar\epsilon \psi,\quad\quad\quad\quad\;\;\;
      \delta \psi=\sqrt 2 \left(i\;\!{\not\!\!{\mathcal D}}\phi-e\, N\phi\right)\epsilon\,,\nonumber\\[2mm]
    \delta \tilde{\phi}&=&\sqrt 2 \bar\epsilon \tilde{\psi},\quad\quad\quad\quad\;\;\;
  \delta \tilde{\psi}=\sqrt 2  \left(i\;\!{\not\!\!{\mathcal D}}\tilde{\phi}+e\,(N+\tilde{m})\tilde{\phi}\right)\epsilon\,,\nonumber\\[1mm]
 \delta A_\mu&=&i(\bar\epsilon\gamma_\mu\lambda-\bar\lambda\gamma_\mu\epsilon),\quad
  \delta \lambda=-\gamma^\mu\epsilon (\partial_\mu N-f_\mu)+i \epsilon \frac{D}{e},
\end{eqnarray}
where
$$f_\mu=-\frac{i}{2}\epsilon_{\mu\alpha\beta}F^{\alpha\beta}\,,\quad
D=e^2\left(|\phi|^2-|\tilde{\phi}|^2-\xi\right),
$$
and $\epsilon=(\epsilon_1,\epsilon_2)$ is a complex spinor. The
corresponding supersymmetry current is
\begin{eqnarray}
\label{susy current} j^\mu&=&\sqrt 2 \left(\!{\not\!\!{\mathcal
D}}\phi^*+ie\, N\phi^*\right)\gamma^\mu\psi +\sqrt 2
\left(\!{\not\!\!{\mathcal D}}\tilde{\phi}^*-ie\,
(N+\tilde{m})\tilde{\phi}^*\right)\gamma^\mu\tilde{\psi}
\nonumber\\[2mm]
&+&\left(i\;\!{\not\!\!{\mathcal \,\partial}}N-i\!{\not\!\!{\mathcal
\,f}}+D\right)\gamma^\mu\lambda\,.
\end{eqnarray}
The centrally extended algebra of the supercharges is discussed
below, in Sect.~\ref{centra}, see Eq.~(\ref{susy algebra}). After
elimination of the auxiliary  $D$ field  via  equation of motion, we
get the following scalar potential:
\begin{equation}
V=\frac{e^2}{2}\,  \left[ \xi - ( \phi^*\,\phi - {\tilde
\phi}^*\,\tilde \phi) \right]^2+ N^2 \phi^* \,\phi +(\tilde m+N)^2
\,{\tilde{\phi}}^*\,\tilde{\phi}\,.
  \label{potential}
\end{equation}
If $\xi$ is positive (and we will assume $\xi> 0$) the theory is in
the Higgs regime and
 supports the BPS-saturated vortices. We will assume $\tilde m$
to be positive too.
 If $\tilde m\neq 0$, the vacuum configuration is as
 follows:
 \beq
 \tilde\phi =0\,,\quad N= 0\,,\quad \phi^*\phi =\xi\,.
 \eeq
 Vortices with
nonvanishing winding number correspond to windings of the $\phi$
field \cite{Penin Rubakov tinyakov troitsky}. The fermionic fields
are set to zero in the classical approximation.

We are interested in static solutions; the relevant part of the
Lagrangian, upon the Bogomol'nyi completion \cite{Bogomolny}, takes the form
\begin{eqnarray}
{\cal L}_{\rm BPS} &=& -\frac{1}{2 e^2}\, B^2 -|{\mathcal
D}^i\phi|^2-\frac{e^2}{2}\,  \left[ \xi - \phi^*\,\phi \right]^2
\nonumber\\[2mm]
&=&
-|{\mathcal D}_+\phi|^2
-\frac{1}{2 e^2}\Big[ B -
e^2(|\phi|^2-\xi)\Big]^2
\nonumber\\[4mm]
&&\,\, -\, \xi B-i \partial_k\left(\epsilon_{k l}\phi^*\,{\mathcal
D}_l\,\phi \right), \label{bpsl}
\end{eqnarray}
where $B=\partial_1\,A_2-\partial_2\,A_1$ is the magnetic field and
${\mathcal D}_+\equiv{\mathcal D}_1+i{\mathcal D}_2$

Since the solution is static we have ${\cal H}=-{\cal L}_{\rm BPS}$.
We will label the fields minimizing ${\cal H}$ by the subscript (or
superscript) $v$. They satisfy the following first-order BPS
equations:
\begin{eqnarray}
B_v - e^2\left( |\phi_v|^2- \xi \right)=0\;,\quad {\mathcal D}^v_+\,
\phi_v=0\,. \label{bps equations}
\end{eqnarray}
The boundary conditions are self-evident. Solutions to these BPS
equations in different homotopy classes are labeled by the winding
number $n$. Needless to say, they are well known. A vortex with the
winding number $n$ has the mass
\begin{eqnarray}
\label{mass of vortex final classical} M_v=2 \pi n \,\xi\,,
\end{eqnarray}
where, at the classical level, the parameter $\xi$ on the right-hand
side is that entering the Lagrangian (\ref{N=2 2+1 dimensions}). At
this level the central charge
\begin{eqnarray}
\label{central charge of vortex classical} | Z_v  |=\xi \int d^2 x B
=2 \pi n\, \xi\, .
\end{eqnarray}

The vortex solution breaks  $1/2$ of supersymmetry. More precisely,
the vortex solution is invariant under the supersymmetry
transformations (\ref{susy transformations}) restricted to
$\epsilon=(0, \epsilon_2)$. In Sect.~\ref{section Quantum
Corrections}  we will show that this residual symmetry between
bosons and fermions is strong enough to preserve the BPS saturation
at the quantum level.

\section{Quantum Corrections}
\label{section Quantum Corrections}

In this section we will calculate quantum corrections to the
Fayet--Iliopoulos parameter, the vortex mass and the central charge,
using the regularization outlined in Sect.~\ref{intro}. We will keep
$\tilde m$ large but finite, taking the limit $\tilde m \to\infty$
at the very end. In order to calculate one-loop corrections to the
classical results we will expand the fields around the background
solutions
\begin{equation}
\label{quantum fluctuations definition}
    \phi=\phi_v+\eta, \quad A_\mu=A^v_\mu+a_\mu
\end{equation}
keeping the terms quadratic in $\eta,\,\,a_\mu$. The fields $\eta$
and $a_\mu$ have the mass\,  $m=e\sqrt{2\,\xi}$, while $\tilde\phi$
and $\tilde\psi$ have the mass $\tilde{m}$.\footnote{The superpartners $\psi$ and
$\lambda$ do not have definite masses; the mass matrix for these
fields can be diagonalized providing us with two diagonal combinations,
$\psi'=\frac{\psi+i\lambda}{\sqrt 2}$ and
$\lambda'=\frac{\psi-i\lambda}{\sqrt 2}$. The latter have masses $e\sqrt {2
\xi}$.
Note that both parameters, $e$ and $\sqrt\xi$, have dimensions
$[m]^{1/2}$.
}

\subsection{Fayet--Iliopoulos parameter at one loop}
\label{fayet}

\begin{figure}[th] \centering
\includegraphics[width=3.5 in]{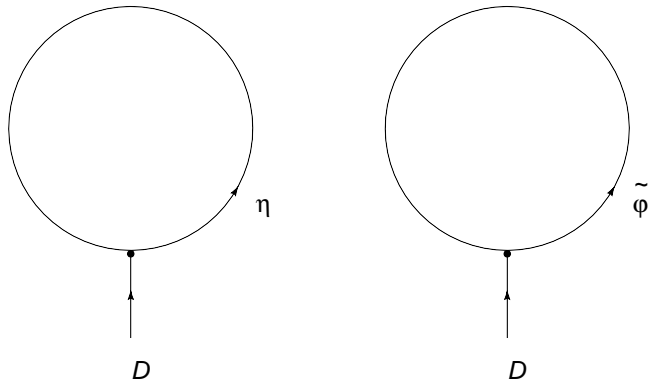}
\caption{\small Tadpole diagrams determining one-loop correction to
$\xi$. } \label{N=2plot}
\end{figure}

As was mentioned, the Fayet--Iliopoulos parameter
receives no corrections if $\tilde m = m$. If  $\tilde m\neq m$,
there is a one-loop quantum correction. The simplest way to compute
the renormalization of $\xi$ is to consider the Lagrangian before
eliminating the auxiliary field $D$, i.e the bosonic part in Eq.
(\ref{N=2 2+1 dimensions}). In this exercise we treat $D$ as a
constant background field. Figure \ref{N=2plot} shows the tadpole
diagrams arising from the couplings
$D( \phi^*\phi- \tilde{\phi}^*\tilde {\phi})$,
which renormalize $\xi$,
\beqn
\label{renorm of ksi}
    \xi_{\rm R} \equiv \xi+\delta\xi
    &=&
    \xi+\int \frac{d^3 k}{(2 \pi)^3}\left(\frac{i}{k^2-\tilde{m}^2}
    -\frac{i}{k^2-m^2}\right)
    \nonumber\\[3mm]
    &=&
    \xi+\frac{m-\tilde m}{4 \pi}\, .
\eeqn We see that $\tilde m$ plays the role of the ultraviolet
cut-off, as was expected. Needless to say, the finite part of the
correction, $m/4\pi$, depends on the definition of the renormalized
FI parameter. In fact, it has an infrared origin (otherwise, odd
powers of $m$ could not have entered). The renormalized FI parameter
is defined as the coefficient in front of the $D$ term in
$\Gamma_{\rm one-loop}$. Here we note a couple of differences
between the result in Eq. (\ref{renorm of ksi}) and the results in
\cite{Rebhan1,vassil}. The first difference is that $\tilde m$,
which represents the linear divergence of $\xi$, is absent in the
previous results since the authors used dimensional and
zeta-function regularization, respectively. Another  difference is
the sign of the $\frac{m }{4 \pi }$ term. The calculation of the
vortex mass renormalization in \cite{Rebhan1,vassil} was phrased as
a counter term calculation; therefore, the result
\cite{Rebhan1,vassil} $\delta\xi=-\frac{m }{4 \pi }$ which
superficially has the sign opposite to that in Eq.~(\ref{renorm of
ksi}) is in full accord with our result and with the central charge
renormalization.

\subsection{Central Charge}
\label{centra}

The nonvanishing (and linearly divergent) correction to $\xi$
implies that the classical central charge in Eq. (\ref{central
charge of vortex classical}) must be corrected too, in accordance
with Eq. (\ref{renorm of ksi}), so that $\xi$ is converted to
$\xi_{\rm R}$ in the central charge. Now we will explain where this
correction comes from.

The centrally extended superalgebra is
\begin{eqnarray}
\label{susy algebra}
\left\{ Q,\,\left(
Q^\dagger\right)\gamma^0\right\} &=& 2\left(P_0\, \gamma^0 +
P_1\,\gamma^1 +P_2\,\gamma^2\right)
\nonumber\\[3mm]
&-& 2\left(P_3\, +\, \xi\,\int d^2x B\right) ,
\end{eqnarray}
where our conventions for the gamma matrices are summarized in
Eq.~(\ref{gamma}) and $P_3$ is the ``momentum" along the reduced
direction,
\begin{equation}
\label{Momentum along 3rd axis}
P_3=-\tilde m \int d^2 x \left( i
\tilde\phi^* \stackrel{\leftrightarrow}{\partial_t}\tilde\phi
+\bar{\tilde\psi}\gamma_0\tilde\psi\right)
\equiv\tilde m \,q\,.
\end{equation}
Here $q$ is the Noether charge of the vortex,
\beq
q=\int d^2 x \tilde J_0\,,\qquad \tilde J_\mu =
-\left( i\tilde\phi^* \stackrel{\leftrightarrow}{\partial_\mu}\tilde\phi
+\bar{\tilde\psi}\gamma_\mu\tilde\psi\right)\,.
\eeq
The current $ \tilde J_\mu$
defines a global U(1) symmetry acting in the regulator sector.
Below we will show that the corresponding charge fractionalizes.
(In the low-energy sector it is related to the occurrence
of the Chern--Simons term after the tilded fermion is integrated out.)

It is rather obvious that the $P_3$ term  is in one-to-one
correspondence with the fact that integrating out massive fermions
in $2+1$ dimensions generates the Chern--Simons term in the
Lagrangian \cite{semenoff,redlich}, which, in turn, makes the vortex
electrically charged \cite{dVS}.  Since our theory is fully
regularized, the superalgebra (\ref{susy algebra}) presents the
exact operator equality in an explicit representation (which is
sometimes elusive in other regularizations.) The second line in
Eq.~(\ref{susy algebra}) is $-2\, Z_v$. Although the coefficient of
the Chern--Simons term in the Lagrangian is dimensionless,
integrating out heavy fermions in the central charge produces a term
which has mass dimension 1. In fact, in Sect. \ref{q} (see also
Appendix) we will calculate the value of the Noether charge $q$ (at
one loop) and will show that $q=-\frac{n}{2}$. Note that for odd $n$
the charge is fractional, a well known phenomenon of charge
fractionalization \cite{goldstone wilczek}.

Assembling two terms in the central charge and using the fact that
$q=-\frac{n}{2}$ we get \beqn \label{renormalized central charge}
   | Z_{n,v} |
   &=&
   2 \,\pi\,n\,\xi+\tilde m \,q=2 \,\pi\,n\,\xi- \frac{\tilde m \,
    n}{2}
    \nonumber\\[1mm]
    &=&
   n\left(    2 \,\pi \,\xi_{\rm R} -\frac{m}{2}\right),
\eeqn where we used Eq. (\ref{renorm of ksi}) to convert $\xi$ into
$\xi_{\rm R}$. The contribution due to $P_3$ comes precisely  in the
combination ensuring that the bare parameter  $\xi$ is converted
into the renormalized $\xi_{\rm R}$.  Equation (\ref{renormalized
central charge}) demonstrates the emergence of the quantum
correction $-m\, n/2$.

\subsection{Renormalization of the vortex mass}
\label{renor}

To calculate the one-loop contribution to the vortex mass,
we expand the Lagrangian (\ref{N=2 2+1 dimensions})
around the background field, in the quadratic order, using the
definitions (\ref{quantum fluctuations definition}). It is
convenient to introduce the following gauge-fixing
term:\,\footnote{This gauge-fixing term is chosen to cancel the terms
 $(\eta^*\, \phi_v)^2$ and  $(\eta\, \phi_v^*)^2$
 originating from the scalar potential
(\ref{potential}) as well as
 the term $\partial_\mu a^\mu ( \eta^*\, \phi_v
- \eta\, \phi_v^*)$ arising from the term ${\mathcal
D}^\mu\phi^*\, {\mathcal D}_\mu \phi$ in Eq. (\ref{bosonic
lagrangian}).}
\begin{equation}
\label{gauge fixing}
    {\cal L}_{\rm gf}=-\frac{1}{2}\Big( \frac{1}{e}\partial_\mu
    a^\mu+ie
    (\phi_v\eta^*-\phi_v^*\eta)\Big)^2\,.
\end{equation}
Note that under this gauge choice, $a_0$ becomes a dynamical field,
and one has to take its loop contribution into account. The
corresponding ghost Lagrangian is
\begin{equation}
\label{ghost}
    {\cal L}_{\rm gh}= \bar c \Big[ -\frac{1}{ e^2}\partial_\mu \partial^\mu-
    (2 |\phi_v|^2+\phi_v\eta^*+\phi_v^*\eta)\Big]c\, ,
\end{equation}
where $\bar c$ and $c$ are spin-zero complex fields with  fermion
statistics. We will drop the last two terms in Eq. (\ref{ghost})
since they show up only in  higher-order corrections. Assembling all
the bosonic contributions, we get  the following bosonic Lagrangian
(at the quadratic order)
\begin{eqnarray}
\label{bosonic fluctuation all}
   L_B^{(2)}
   \!\!\!
   &=&{\mathcal L}^{(2)}_{\rm gf}+{\mathcal L}^{(2)}_{B}+
   {\mathcal L}_{\rm gh}^{(2)}
   \nonumber\\[2mm]
   &=&|{\mathcal D}^v _\mu \eta|^2-e^2\left(3
    |\phi_v|^2-\xi\right)|\eta|^2
       \nonumber\\[2mm]
     &+&
     \frac{1}{2 e^2}(\partial_\mu
    a_m)^2
-
     |\phi_v|^2 a_m^2-2i a^m\left(\eta^*{\mathcal D}^v_m\phi_v-
     \eta{\mathcal D}^v_m\phi_v^*\right)
          \nonumber\\[2mm]
     &+&
    |{\mathcal D}^v_\mu\tilde{\phi}|^2+\Big[ e^2\,(|\phi_v|^2-\xi^2)-\tilde m^2
    \Big]
    |\tilde\phi|^2
     \nonumber\\[2mm]
       &-& \!\!
       \frac{1}{2 e^2}(\partial_\mu
    a_0)^2+|\phi_v|^2 a_0^2+\frac{1}{2 e^2}(\partial_\mu N)^2-N^2  |\phi_v|^2
     \nonumber\\[2mm]
     &+&  \bar{c}\left( -\frac{1}{ e^2}\partial_\mu \partial^\mu-
    2 |\phi_v|^2\right)c\,,
    \rule{0mm}{6mm}
\end{eqnarray}
where $\mu=0,\,1,\,2$  and  $m=1,\,2$ (the fields $\eta$ and $a$ are
defined in Eq.~(\ref{quantum fluctuations definition})). The last two
lines in Eq. (\ref{bosonic fluctuation all}) include one complex
scalar fields with the fermion statistics and two
real scalar fields with the boson statistics,
satisfying the same equations of motion.  If we impose the same
boundary conditions on the fields $a_0$, $N$ , $\bar c$ and $c$,
(and we do),
they produce the same determinants, and their contributions to the
vortex mass cancel each other~\cite{vassil}. With this observation
in mind, we will drop this line in what follows.

The transverse
components of the gauge field, $a_1$ and $a_2$, can be combined into
 complex fields by defining
 \begin{equation}
 a^{\pm}=\frac{a^1\pm i a^2}{\sqrt{2} e}.
 \end{equation}
By the same token, we define ${\mathcal D}^v_\pm={\mathcal D}^v_1\pm
\,i{\mathcal D}^v_2$. With these definitions Eq. (\ref{bosonic
fluctuation all}) can be rewritten as follows:
\begin{eqnarray}
\label{bosonic fluctuation all simplified}
   L_B^{(2)}&=&|{\mathcal D}^v _\mu \eta|^2-e^2(3
    |\phi_v|^2-\xi)|\eta|^2
    \nonumber\\[2mm]
    &+&
    \partial_\mu
    a^+\partial^\mu a^-
     -2 e^2 |\phi_v|^2 a^+ a^- -\sqrt 2 i e \left(\eta^*a^+{\mathcal D}^v_-\phi_v-\eta a^-{\mathcal D}^v_+\phi_v^*
    \right)\nonumber\\[2mm]
    &+&
     |{\mathcal D}^v_\mu\tilde{\phi}|^2+\left( e^2\,(|\phi_v|^2-\xi^2)-\tilde m^2
    \right)|\tilde\phi|^2
    \,.
\end{eqnarray}
Note that, at the quadratic order, the tilded bosonic sector is
decoupled from the fluctuations of the nontilded one, i.e.
$\tilde{\phi}$ is coupled to the background fields only. (We will
soon see that the same decoupling occurs for the fermionic sector.)
This allows us to consider the contributions of tilded and untilded
fields separately.

 \subsubsection{One-loop contribution from the untilded sector}
 \label{onelo}
In the first part of this subsection we will compute the classical
Hamiltonian (density) of the fluctuations. In the second part we
will quantize the Hamiltonian  by imposing canonical
(anti)commutation relations. Finally we will compute the sum of the
energies, which turns out to be vanishing. We first start with the
bosonic Hamiltonian corresponding to the untilded part of the
Lagrangian (\ref{bosonic fluctuation all simplified}), which can be
written in the matrix form,
\begin{eqnarray}
\label{H operator}
  {\cal H}_B^{(2)}&=&
  \left(
  \begin{array}{cc} \dot{\eta}, &
  i\,\dot{a}_+ \\\end{array}
\right)^* \left(
  \begin{array}{c}
   \dot{ \eta} \\
    i\,\dot{a_+} \\
  \end{array}
\right)
  +\left(
  \begin{array}{cc}
\eta, & i\, a_+  \\
  \end{array}
\right)^* D^2_B\left(
  \begin{array}{c}
    \eta \\
    i\, a_+ \\
  \end{array}
\right),
\end{eqnarray}
where we defined the quadratic bosonic operator
\begin{eqnarray}
\label{bosonic operator square}
  D^2_B &=& \left(
          \begin{array}{ccc}
            - \left(\mathcal{D}^{v\,}_k\right)^2 +e^2(3
    |\phi_v|^2-\xi)&  \sqrt 2 e\mathcal{D}^v_-\phi_v \\[5mm]
             \sqrt 2 e(\mathcal{D}^v_-\phi_v )^*& -\partial_k^2+2 e^2 |\phi_v|^2
          \end{array}
        \right).
\end{eqnarray}
Eq. (\ref{H operator}) gives the classical Hamiltonian for the
bosonic fields.   The fermionic Lagrangian (\ref{bosonic
lagrangian}) is already quadratic in the fermionic fields.  Setting
the bosonic fields to their background values gives the following
quadratic Lagrangian for the untilded fermionic fields:
\begin{eqnarray}
{\mathcal L}^{(2)}_{F}&=&\frac{1}{e^2}\bar\lambda
\,i\,\!{\not\!\partial}\,\lambda  + \bar\psi
\,i\,\!{\not\!\!{\mathcal D}}\,\psi +i
\sqrt{2}\left[\left(\bar\lambda\,\psi\phi^*-\bar\psi \lambda\,
\phi\right)\right] \label{fermionic fluctuation}.
\end{eqnarray}
We choose the following set of $\gamma$ matrices:
\begin{equation}
\label{gamma}
    \gamma^0=\sigma_3,\;\; \;
    \gamma^1=i\,\sigma_2,\;\; \;
    \gamma^2=i\,\sigma_1\,.
\end{equation}
With the chosen representation of  $\gamma$ matrices the Hamiltonian
corresponding to the Lagrangian (\ref{fermionic fluctuation}) reads
\begin{eqnarray}
  {\cal H}_F^{(2)} \! &=&
  \!\!
  -i\left(
                 \begin{array}{c}
                   \psi_1 \\ \psi_2  \\\lambda_1/e\\ \lambda_2/e  \\
                 \end{array}
               \right)^\dagger
\left(
  \begin{array}{cccc}
    0 &  {\cal D}^v_+ &   -\sqrt 2 e \phi_v  & 0 \\
     {\cal D}^v_- & 0 & 0 &  \sqrt 2 e \phi_v  \\
     \sqrt 2 e \phi_v^* & 0 & 0 & \partial_+ \\
    0 & -  \sqrt 2 e \phi_v^*  & \partial_- & 0 \\
  \end{array}
\right)\left(
         \begin{array}{c}
          \psi_1 \\
           \psi_2 \\
           \lambda_1/e\\
           \lambda_2/e \\
         \end{array}
       \right)\nonumber\\[6mm]
                  &=&
                   \left(
    \begin{array}{c}
      U \\
      V
    \end{array}
  \right)^\dagger\left(
             \begin{array}{cc}
               0 & - i D_F \\
                i D^{\dagger}_F & 0 \\
             \end{array}
           \right)\left(
                    \begin{array}{c}
                      U \\
                      V \\
                    \end{array}
                  \right),
\end{eqnarray}
where we regrouped the components of $\lambda$ and $\psi$, \beq
\left\{
    \begin{array}{cc}
      U; &
      V \\
    \end{array}
  \right\}
   =\left\{
         \begin{array}{cccc}
            ( \psi_1, &
           \lambda_2/e)\,; &
             (\psi_2, &
           \lambda_1/e) \\
         \end{array}
       \right\},
       \eeq
and defined the fermionic operator,
\begin{eqnarray}
D_F\equiv\left(
  \begin{array}{cc}
    {\cal D}^v_+& -\sqrt 2 e \phi_v\\[3mm]
    -\sqrt 2 e \phi_v^*& \partial_-
  \end{array}
\right),\quad D_F^\dagger=\left(
  \begin{array}{cc}
    -{\cal D}^v_-& -\sqrt 2 e \phi_v\\[3mm]
    -\sqrt 2 e \phi_v^*& -\partial_+
  \end{array}
\right).
\end{eqnarray}
Supersymmetry of the Lagrangian reveals itself when we calculate the
following quadratic fermionic operator:
\begin{eqnarray}
\label{fermion operator square} D^{\dagger}_F D_F=\left(
          \begin{array}{ccc}
            - \left( {\mathcal D}^{v}_k\right)^2 +e^2(3
    |\phi_v|^2-\xi)&  \sqrt 2 e {\cal D}^v_-\phi_v \\[4mm]
             \sqrt 2 e ({\mathcal D}^v_-\phi_v )^*& -\partial_k^2+2  e^2 |\phi_v|^2
          \end{array}
        \right),
\end{eqnarray}
which coincides with $D^2_B$ defined in Eq. (\ref{bosonic operator
square}), \beq D_B^2 = D^{\dagger}_F D_F\,. \label{dbdf} \eeq By
virtue of this identification we rewrite the full Hamiltonian for
untilded fields in terms of the operators $D_F$ and $D^\dagger_F$,
\begin{eqnarray}
\label{H full}
  {\cal H}^{(2)}&=&
  \left(
  \begin{array}{cc} \dot{\eta}, &
  i\,\dot{a}_+ \\
  \end{array}
\right)^* \left(
  \begin{array}{c}
   \dot{ \eta} \\
   i \,\dot{a_+} \\
  \end{array}
\right)+\left(
  \begin{array}{cc}
\eta, & i\, a_+  \\
  \end{array}
\right)^* D_F^\dagger D_F\left(
  \begin{array}{c}
    \eta \\
   i\, a_+ \\
  \end{array}
\right)
  \nonumber\\[3mm]
   &-& i U^\dagger \,  D_F\,  V+ i V^\dagger\,D^{\dagger}_F\, U.
\end{eqnarray}
To quantize the Hamiltonian (\ref{H full}) we will follow methods
worked out long ago (e.g. \cite{yamagishi}). First, we impose
boundary conditions which are compatible with the residual
supersymmetry. We place the system into a spherical two-dimensional
``box" of radius $R$, with the assumption that $R$ is much larger
than any length scale in the model at hand. To ensure that the
energy associated with the boundary vanishes, we require all the
fields to vanish at $r=R$. This condition does not break  the
residual supersymmetry since it is compatible with the
transformations defined in Eq. (\ref{susy transformations}). Then we
expand the fields in Eq. (\ref{H full}) in eigenmodes of the
operators $D^2_B$ and the associated operator $  D^{2\,\,'}_B$
defined as follows:
\begin{equation}
\label{bosonic operator reverse order} D^{2\,\,'}_B= D_F
D^\dagger_F.
\end{equation}
The eigenvalue equations for these operators are
 \begin{equation}
 \label{eigenvalue equation}
    D^2_B\, \xi_{n,\sigma}\equiv w_n^2\,  \xi_{n,\sigma}, \quad D^{2\,\,'}_B\, {\xi}'_{n,\sigma}\equiv{w}_n^2
    \,{\xi}'_{n,\sigma}.
\end{equation}
The eigenvalues for both operators are the same: the eigenfunctions
can be related to each other by
 \begin{equation}
 \label{bosonic operator reverse order}
  {\xi}'_{n,\sigma}=\frac{1}{w_n} D_F \, \xi_{n,\sigma}\,,
  \qquad   \xi_{n,\sigma}=\frac{1}{w_n} D^\dagger_F\,
  {\xi}'_{n,\sigma}
  \,.
\end{equation}
For each $w_n^2$ there are two independent solutions, which are
labeled by subscript $\sigma$. The above statement excludes the zero
modes, $w_n=0$, which occur only in one of these operators, namely
$D_B^2$, reflecting the translational invariance in the problem at
hand. Usually, they are referred to as translational. Their fermion
counterparts, the zero modes of $D_F$, are supertranslational modes.
$D_F^\dagger$ has no zero modes.

The eigenfunctions $\xi_{n, \sigma}$ form an orthonormal and
complete basis, in which we expand the fields in Eq. (\ref{H full})
\begin{eqnarray}
\label{eigenvalue expansion} &&
 \left(\begin{array}{c}
    \eta (t,{\bf x})\\[2mm]
    i \,a_+(t,{\bf x})
  \end{array}
\right)= \sum_{\begin{array}{ll}
n\neq 0\\
\sigma=1,2
\end{array}
} a_{n,\sigma}(t)\, \xi_{n,\sigma}({\bf x})\, ,
\nonumber\\[3mm]
&& V(t,{\bf x})=\sum_{\begin{array}{ll}
n\neq 0\\
\sigma=1,2
\end{array}} v_{n,\sigma}(t) \, \xi_{n,\sigma}({\bf x}),\nonumber\\[3mm]
&&
 U(t,{\bf x})=\sum_{\begin{array}{ll}
n\neq 0\\
\sigma=1,2
\end{array}}
u_{n,\sigma}(t)\,  {\xi_{n,\sigma}'}({\bf x})\,.
\end{eqnarray}
Note that the zero modes do not enter in the expansion
(\ref{eigenvalue expansion}), nor do they appear in the Hamiltonian
(\ref{H full}). For nonzero modes the ratio of the bosonic to
fermionic modes is 1:2, i.e. we have two complex expansion
coefficients $a_{n,\sigma}(t)$ for bosons and four complex expansion
coefficients $v_{n,\sigma}(t)$ and $u_{n,\sigma}(t)$ for fermions,
for each value of $w_n^2$. As we will see below, this is precisely
what is needed for cancelation. Let us note in passing that for zero
modes the ratio is 1:1. We have one complex bosonic modulus and one
fermionic.

Using the above mode decompositions in Eq. (\ref{H full}),
 we arrive at an infinite set of oscillators,\footnote{To be
 accurate we should note that here we use integration by parts in the last term,
 which means that Eq.~(\ref{Hamiltonian eigenvalue expansion}) is valid up to a full spatial derivative.}
\begin{eqnarray}
\label{Hamiltonian eigenvalue expansion}
 {\mathcal H}^{(2)}
&=& \sum_{\begin{array}{ll}
n, n'\neq 0\\
\sigma , \sigma '
\end{array}
} \Big( \dot{a}^*_{n,\sigma} \dot{a}_{n',\sigma '}\,
\xi_{n,\sigma}^\dagger\xi_{n',\sigma '}  + {a}^*_{n,\sigma}
{a}_{n',\sigma '}\, \xi_{n,\sigma}^\dagger D_F^\dagger D_F
\xi_{n',\sigma '}
\nonumber\\[3mm]
&-& i \, \frac{1}{w_{n'}} \,  v_{n,\sigma}^* u_{n',\sigma
'}\,\xi_{n,\sigma}^\dagger \,  D_F^\dagger D_F\, \xi_{n',\sigma '}
 +
i \,w_n\,  u^*_{n,\sigma} v_{n',\sigma '} \, \xi_{n,\sigma }^\dagger
\xi_{n',\sigma '}\Big)\,.
\end{eqnarray}
Now,  for each oscillator,  the coefficients $a$, $\dot a$, $v$, $u$
and their complex conjugated must be represented as linear
combinations of the corresponding creation and annihilation operator
subject to the standard (anti)commutation relations. This procedure
parallels that discussed in detail in Ref.~\cite{SVV}. The only
difference is that in \cite{SVV} for each mode one has an oscillator
for one real degree of freedom, while in  the case at hand we deal
with a complex degree of freedom which is equivalent to two real
degrees of freedom. We will not dwell on details referring the
reader to Ref.~\cite{SVV}. Imposing the appropriate
(anti)commutation relations on the creation and annihilation
operators, we get for expectation values of bilinears in the vortex ground
state
\beqn
\label{bilinears}
\langle a_{n,\sigma}^*
a_{n',\sigma '}\rangle_{\rm vor} &=& \frac{1}{2
w_n}\delta_{nn'}\delta_{\sigma \sigma '},
\qquad\langle\dot{a}_{n,\sigma}^* \dot{a}_{n',\sigma '}\rangle_{\rm
vor} =\frac{w_n}{2} \delta_{nn'}\delta_{\sigma \sigma '},
\nonumber\\[2mm]
 \langle u_{n,\sigma}^*
v_{n',\sigma '}\rangle_{\rm vor} &=&
 \frac{i}{2} \delta_{nn'}\delta_{\sigma \sigma '},\qquad
\langle v_{n,\sigma}^* u_{n',\sigma '}\rangle_{\rm vor} =
 -\frac{i}{2} \delta_{nn'}\delta_{\sigma \sigma '}\,,
\eeqn
where the angular brackets mark the vortex expectation value.
Expectation values of all other bilinears vanish.
 If we substitute
these results in Eq.~(\ref{Hamiltonian eigenvalue expansion}) we
immediately see that the one-loop correction in the untilded sector
vanishes locally, i.e. in the Hamiltonian density. Needless to say,
it vanishes in the integral $\int d^2x \, {\mathcal H}^{(2)}$ too.

Thus, we demonstrated the cancelation of the bosonic and fermionic
contributions mode by mode, for each given $n$. This vanishing
result shows that the vortex mass receives no correction from the
untilded sector. If we did not have the $\tilde\Phi$ multiplet, this
would be the final answer. However, the theory per se is ill-defined
without the tilded sector.

From Eq. (\ref{renorm of ksi}) we see that in the absence of
$\tilde\phi$, the FI parameter $\xi$ would be linearly divergent at
one loop. With $\tilde\phi$ included, the theory is regularized;
cancelation of loops in Fig. \ref{N=2plot} takes place. The linear
divergence is replaced by the linear dependence of $\xi$ on $\tilde
m$. The latter parameter  is kept large, but finite till the very
end. It is only natural that the linear dependence of $M_{v, \rm R}$
on $\tilde m$ will be provided by the tilded sector contribution
(Sect.~\ref{theti}).

\subsubsection{The tilded sector (regulator) contribution in $M_v$}
 \label{theti}

The Lagrangian for the tilded sector is
\begin{eqnarray}
\label{tilde lagrangian}
   \tilde{\mathcal L}^{(2)}&=&|{\mathcal D}^v_\mu\tilde{\phi}|^2+
   \Big( e^2\,(|\phi_v|^2-\xi^2)-\tilde m^2    \Big)
    |\tilde\phi|^2+\bar{\tilde\psi }\,i\,\!{\not\!\!{\mathcal
D}}\,\tilde\psi - \tilde m\, \bar{\tilde\psi}\,\tilde\psi.
\end{eqnarray}
The corresponding Hamiltonian density then takes the form
\begin{eqnarray}
\label{tilde hamiltonian}
   \tilde{{\cal H}}^{(2)}&=& |\dot{\tilde\phi}|^2+\,{\tilde\phi}^*(-{\mathcal D}^{v}_
   +{\mathcal D}^{v}_-+
             \tilde{m}^2){\tilde\phi}
             \nonumber\\[5mm]
             &+&
               \left(
            \begin{array}{cc}
              \tilde\psi_1 &  \tilde\psi_2
            \end{array}
          \right)^*
      \left(
            \begin{array}{cc}
              \tilde{m} &-i  {\mathcal D}^v_+\\[2mm]
              - i {\cal D}^v_-& -\tilde{m}
            \end{array}
          \right) \left(
                    \begin{array}{c}
                       \tilde\psi_1 \\
                       \tilde\psi_2 \\
                    \end{array}
                  \right),
\end{eqnarray}
where ${\mathcal D}^v_\pm={\mathcal D}^v_1 \pm i{\mathcal D}^v_2$
and we used Eq.~(\ref{bps equations}). For what follows it is
important to know that the operator ${\mathcal D}^v_+{\mathcal
D}^v_-$ has no zero modes.

If we denote the eigenvalues of the bosonic operator \beq -{\mathcal
D}^v_+{\mathcal D}^v_- +\tilde m^2 \eeq by $\Delta $ ($\Delta $ is
strictly larger than $\tilde m^2$), for each given $\Delta $ we have
two eigenmodes of the associated fermion equation \beq
 \left(
            \begin{array}{cc}
              \tilde{m} &-i  {\mathcal D}^v_+\\[2mm]
              - i {\cal D}^v_-& -\tilde{m}
            \end{array}
          \right) \left(
                    \begin{array}{c}
                       \tilde\psi_1 \\
                       \tilde\psi_2 \\
                    \end{array}
                  \right) =
                  \pm \sqrt\Delta \, \left(
                    \begin{array}{c}
                       \tilde\psi_1 \\
                       \tilde\psi_2 \\
                    \end{array}
                  \right).
                  \label{45}
\eeq The eigenfunctions have the following structure. If
$\tilde\psi_1$ is the normalized eigenfunction of the operator
$-{\mathcal D}^v_+{\mathcal D}^v_-$, then $\tilde\psi_2$ is the
corresponding eigenfunction of the conjugated operator $-{\mathcal
D}^v_-{\mathcal D}^v_+$ times
$$
\sqrt{\frac{\pm \sqrt\Delta -\tilde m}{\pm \sqrt\Delta +\tilde m} }
$$
depending on the sign in the eigenvalue equation (\ref{45}). Thus,
for each complex boson mode with the eigenvalue $\Delta$ we have two
complex fermion modes with the eigenvalues $\pm\sqrt\Delta$. This
balance of modes guarantees that the corresponding quantum
corrections to $M_v$ vanish.

This is not the end of the story, however. There is {\em one}
additional (complex) fermion mode with $ \Delta$ exactly equal to
$\tilde m^2$. (The above statement refers to the elementary vortex
with the unit winding number. Generalization to higher winding
numbers is straightforward.) Let us focus on this unbalanced mode
which will be solely responsible for the contribution of the tilded
sector in $M_v$.

From Eqs. (\ref{45}) and (\ref{bps equations}) it is clear that this
fermion mode has the form \beq \left(
                    \begin{array}{c}
                       0\\
                       \tilde\psi_2^{(0) }\\
                    \end{array}
                  \right),
\eeq where the eigenvalue on the right-hand side of Eq. (\ref{45})
is $-\tilde m$. This gives rise to the following contribution in the
energy density:

\begin{equation}
\label{zero mode energy} {\cal E}^{(0)} =-\tilde m\,
          (\tilde{\psi} ^{(0)} _2)^*\, \tilde{\psi} ^{(0)}_2\,.
\end{equation}
We proceed to quantization in the standard manner. To this end we
represent
\beq
\label{zero mode operator form}
 \tilde{\psi}^{(0)}_2 =
{\alpha }^\dagger(t) \,  \varphi({\bf x})\,,
 \eeq
where $\varphi({\bf x})$ is the normalized $c$-numerical part of the
zero mode  while $\alpha^\dagger$ is the operator part with the
appropriate anticommutation relation implying
\begin{equation}
\label{zero mode exp quant} \langle {\alpha }{\alpha
}^\dagger\rangle  =\frac{1}{2}\, .
\end{equation}
Now, the contribution of the tilded sector to $M_v$ obviously
reduces to
\begin{equation}
\label{zero mode energy total} \delta M\equiv\int d^2x\,
\langle{\mathcal E^{(0)}}\rangle =-\tilde m \,\langle {\alpha
}{\alpha }^\dagger\rangle =-\frac{\tilde m}{2} \,.
\end{equation}
Equation (\ref{zero mode energy total}) gives the only nonvanishing
quantum
 correction,
\begin{equation}
\label{mass renormalizedd} M_{\rm R} \equiv M+\delta M=2 \pi\,
\xi-\frac{\tilde m }{2}=2 \pi\, \xi_{\rm R} -\frac{ m}{2}\, ,
\end{equation}
where we again used Eq. (\ref{renorm of ksi}) to convert $\xi$ to
$\xi_{\rm R}$. Comparing this result with the renormalization of the
central charge in Eq. (\ref{renormalized central charge}), we
conclude that the BPS saturation does indeed hold at the quantum
level.

\subsection{Higher orders}
\label{hior}

Let us discuss now what changes as we pass to higher orders of
perturbation theory. Returning to Sect.~\ref{centra}
and, in particular, to Eq.~(\ref{Momentum along 3rd axis}),
it is not difficult to understand that
the relation $Z = 2\pi\xi -\frac{1}{2}\tilde m$
(for the elementary vortex) remains {\em exact} to all orders.
Indeed, $q$ is half-integer and the relation $q=-\frac{1}{2}$
for the elementary vortex
cannot receive corrections\,\footnote{A simple dimensional
 analysis shows that
perturbative corrections run in powers of $e/\sqrt\xi$.}
 in $e/\sqrt\xi$.
 If we define $\tilde\xi$ as
 \beq
 \tilde\xi = \xi -\frac{\tilde m}{4\pi}\,,
 \eeq
where $\xi$ and $\tilde m$
are bare parameters,
then the statement that
\beq
M_v = Z = 2\pi\tilde\xi
\label{54ors}
\eeq
is valid to all orders. The term $\tilde m$ comes from
the ultraviolet, and, therefore,  it is natural to refer
to $\tilde \xi$  as to an
``effective ultraviolet parameter."
Equation (\ref{54ors}) is akin to the NSVZ theorem
for the gauge coupling renormalization in four dimensions
\cite{nsvz}: being expressed of terms of the ultraviolet (bare) parameters
the gauge coupling renormalization is limited to one loop
(see also the second paper in \cite{noxi}).

Corrections in powers of $e/\sqrt\xi$ arise if we decide to express the
result in terms of $\xi_R$, a parameter defined in the infrared;
the expression of $\xi_R$ in terms of $\xi$
does contain an infrared contribution (otherwise, odd
powers of $e$ could not have entered,
see Sect.~\ref{fayet}). Generalizing the arguments of \cite{noxi}
we can write, instead of (\ref{renorm of ksi})
\beq
\tilde\xi = \xi_R\left(1- \frac{1}{2\sqrt{2}\,\pi}\,\frac{e}{\sqrt\xi_R}
\right)\,.
\label{54orsp}
\eeq
Equations (\ref{54ors}) and (\ref{54orsp})
assembled together present a
perturbatively exact result for $M_v = Z$.

\section{Calculation of the Noether charge $q$}
\label{q}

\begin{figure}[th] \centering
\includegraphics[width=0.8 in]{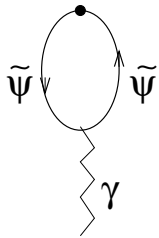}
\caption{\small Calculation of the vortex Noether charge. We trace
the term linear in the momentum $k$ of the produced photon (assuming
$k$ to be small). Terms of the zeroth order vanish because of the
gauge invariance. Quadratic and higher order terms in $k$ are
irrelevant since they are suppressed by powers of $1/\tilde{m}$.}
\label{fig2} \end{figure}

In Sect.~\ref{centra}  we used the fact that the Noether U(1) charge
of the elementary vortex is $-1/2$. The Noether charge is saturated
by the fermion term in Eq.~(\ref{Momentum along 3rd axis}), \beq q=-
\int d^2 x\, \bar{\tilde \psi}\gamma^0\tilde\psi\,. \label{cnc} \eeq
Here we will explore this issue in more detail. The vortex Noether
charge can be calculated in a number of ways. The most
straightforward calculation is that of the Feynman diagram depicted
in Fig.~\ref{fig2}, using the background field expansion. This
expansion is justified because the background photon field is small
compared to the value of $\tilde m$ (in the very end we want to tend
$\tilde m$ to infinity).
 For our purposes it is sufficient to limit
ourselves to the leading term (proportional to $F_{\alpha\beta}$).
Using $\gamma^\mu$ in the upper vertex in Fig.~\ref{fig2} (denoted
by the closed circle) we get the Noether current in  the background
field in the form \beq \label{cs term}\bar{\tilde
\psi}\gamma^\mu\tilde\psi\longrightarrow \tilde{m}\, F_{\alpha\beta}
\,\epsilon^{\mu\alpha\beta}\, \int \frac{d^3p}{(2\pi)^3}\,
\frac{1}{(p^2 +\tilde{m}^2)^2}= \frac{1}{8\pi}\,  F_{\alpha\beta}
\,\epsilon^{\mu\alpha\beta}\,. \eeq The current in  (\ref {cs term})
couples to the gauge field $A_\mu$, giving a term of the form $A_\mu
F_{\alpha\beta} \,\epsilon^{\mu\alpha\beta}$, which is nothing but
the Chern--Simons term.  Now, if we set $\mu=0$ in (\ref {cs term})
and invoke the standard value of the magnetic flux,
$$
\int d^2 x\, B = 2\pi\,,
$$
we immediately get \beq\label{charge result0} \langle q \rangle =-
1/2. \eeq

\section{Conclusion}
\label{concl}

In this paper we showed that the mass and the central charge of the
${\cal N}=2$ vortices in $2+1$ dimensions, being expressed in terms of $\xi_R$,
 get a quantum
correction  $-m \, n/2$ where $m$ is the mass of the charged bosons
(fermions) and $n$ is the winding number of the vortex. The
equality of the corrections to the
vortex mass/central charge shows that the BPS saturation persists at
the quantum level. Our result is in agreement with the previous
ones \cite{vassil,RebhanPN}.

New elements of our work (compared to
\cite{vassil} and \cite{RebhanPN}) are as follows. We use a more straightforward
and physically transparent
regularization scheme which captures linearly divergent terms
invisible in the regularization methods used in the
previous papers. In our scheme we have a massive
regulator multiplet acting  in
loops as an ultraviolet cutoff. In the limit of
infinitely large regulator mass, regulator's role is
taken over by the Chern--Simons term. We establish a contact between
one-loop calculations and the general
operator expression for the central charge (obtained within
the same regularization scheme). Analyzing both, in a single package,
we are able to reveal a simple physical interpretation behind the occurrence
of
the $-m \, n/2$ shift, and obtain all-order results (Sect.~\ref{hior}).

\section*{Acknowledgments}

We would like to thank A. Vainshtein, M. Voloshin, A. S. Goldhaber, A.~Rebhan, P.
van Nieuwenhuizen, and D. V. Vassilevich for useful discussions. This
work is supported in part by DOE Grant DE-FG02-94ER-40823. M.S. is
supported in part by {\em Chaire Internationalle de Recherche Blaise
Pascal} de l'Etat et de la R\'{e}goin d'Ille-de-France,
g\'{e}r\'{e}e par la Fondation de l'Ecole Normale Sup\'{e}rieure.

\vspace{0.5cm}

\section*{Appendix}

\addcontentsline{toc}{section}{Appendix }

\renewcommand{\theequation}{A.\arabic{equation}}
\setcounter{equation}{0}

 \renewcommand{\thesubsection}{A.\arabic{subsection}}
\setcounter{subsection}{0}

It is instructive to illustrate calculations of the charge $q$ by
inspecting the fermion mode decomposition  discussed in
Sect.~\ref{theti}. It is important to note that the mode
decomposition in Sect.~\ref{theti} is {\em not} the canonical expansion.
A similar charge calculation by virtue of the  canonical expansion was first
performed in \cite{jackiw rebbi}. We will discuss both methods.

First, we expand the fields $\tilde\psi_1$ and $\tilde\psi_2$
in terms of the eigenfunctions of the operators $-\mathcal{D}^v_+
\mathcal{D}^v_-$ and $-\mathcal{D}^v_- \mathcal{D}^v_+$,
namely, in
$\eta_{n,\,\sigma}$ and ${\eta'}_{n,\,\sigma}$, respectively:
\begin{eqnarray}
\label{fermion expansion with zero modes}
\tilde\psi_1&=& \sum_{\begin{array}{ll}
n\neq 0\\
\sigma=1,2
\end{array}}
v_{n,\sigma}(t)\, {\eta}_{n,\,\sigma}({\bf x})
\,,
\nonumber\\[2mm]
\tilde\psi_2&=&\tilde\psi^{(0)}_2+\sum_{\begin{array}{ll}
n\neq 0\\
\sigma=1,2
\end{array}}
u_{n,\sigma}(t)\, {\eta}'_{n,\,\sigma}({\bf x}),
\end{eqnarray}
where $\sigma$ labels two independent solutions corresponding to the
same eigenvalue, and   $\tilde\psi^{(0)}_2$  is the zero mode
defined in Eq. (\ref{zero mode operator form}). The nonvanishing
bilinears constructed from $u_{n,\sigma}(t)$ and $v_{n,\sigma}(t)$
are given in Eq. (\ref{bilinears}). With the  expansion in Eq.
(\ref{fermion expansion with zero modes}) in hands, it is easy to see that
the only nonvanishing contribution to $q$ comes from the zero mode
of the operator ${\mathcal D}_+^v$. This statement is a consequence of
the following expansion of $q$:
\begin{eqnarray}
\label{charge expansion}
    \langle q\rangle&=&-\int
    d^2x\langle\tilde\psi^\dagger\tilde\psi\rangle
    \nonumber\\[2mm]
&=& -\langle \alpha\alpha^\dagger\rangle-
\!\!\!\sum_{\begin{array}{ll}
n\neq 0\\
\sigma=1,2
\end{array}}
\langle u_{n,\sigma}^*u_{n,\sigma}+v_{n,\sigma}^*v_{n,\sigma}\rangle\,.
\end{eqnarray}
Using  Eqs. (\ref{bilinears}) and (\ref{zero mode exp quant})
we get
\begin{equation}
\label{charge result1}
    \langle q\rangle=-{1}/{2},
\end{equation}
in perfect agreement with the previous result
(\ref{charge result0}).

(The fact that  $q=-1/2$ on the vortex is in one-to-one
correspondence with the fact that integrating out the massive
fermion $\tilde\psi$ we  generate the Chern--Simons term with
$\kappa=\frac{e}{4 \pi}$ \cite{redlich}. It is well known that
selfdual $n$-vortices with the Chern--Simons term have charge
$q=-\frac{2 \pi n\kappa}{e}=-\frac{n}{2}$ where $n$ is the winding
number \cite{Paul Khare,dVS}.)

We can carry out a slightly different calculation
of  the $q$ charge by expanding the tilded fermion field in the
canonical basis. However, we should remember that, generally
speaking, the U(1) charge of the vacuum is infinite in the absence
of proper regularization.\,\footnote{This infinity
does not show up in  Eq. (\ref{charge result1}) because a
regularized definition  (\ref{charge definition}) of the $q$ charge is built
in in the  expansion coefficients.} The same ``vacuum" infinity then shows up in
$q$. In fact, we are interested in the difference between the values of $q$
on the vortex and in the vacuum.

 This problem is automatically solved if, instead of the
charge  $-\int d^2 x\, \tilde\psi^\dagger\tilde\psi$,  one uses the
following definition:
\begin{equation}
\label{charge definition}
    q=- \frac{1}{2}\,
    \int d^2 x \left(\tilde\psi^\dagger\tilde\psi-\tilde\psi_c^\dagger\tilde\psi_c\right),
\end{equation}
where $\tilde\psi_c=-i(\tilde\psi^\dagger\gamma_2)^T $ is the
charge-conjugated fermion field. We now expand the fermionic field
$\tilde\psi$ in the canonical basis,
\begin{equation}
\label{zero mode full expansion}
    \tilde\psi= {a_0}^\dagger \, \left(
      \begin{array}{c}
        0 \\
        \varphi_0 \\
      \end{array}
    \right)
    +\sum_{\begin{array}{ll}
n\neq 0\\
\sigma=1,2
\end{array}} \left(
 e^{-i w_n t}\frac{a_{n,\sigma}}{\sqrt 2} \varphi_{n,\sigma}+
     e^{i w_n t}\frac{b_{n,\sigma}^\dagger}
     {\sqrt 2} \varphi^*_{n,\sigma}    \right),
\end{equation}
where $\varphi_{n,\sigma}$ are the energy eigenfunctions of the
fermionic Hamiltonian with the eigenvalues $w_n$. The operators $a_{0}$,
$a_{n,\sigma}$ and $b_{n,\sigma}$ obey the canonical anticommutation
relations\,\footnote{Needless to say, all other anticommutators,
not indicated in (\ref{canonical commutations}),
vanish.}
\begin{equation}
\label{canonical commutations}
\{a_{0},a_{0}^\dagger\}=1,\;\;\{a_{n,\sigma},a_{n',\sigma'}^\dagger\}=\delta_{n,n'}\delta_{\sigma,\sigma'}\,,\;\;\{b_{n,\sigma},b_{n',\sigma'}^\dagger\}=\delta_{n,n'}\delta_{\sigma,\sigma'}.
\end{equation}
The operators $a_{n,\sigma}$ and $b_{n,\sigma}^\dagger$ are
the annihilation and creation operators associated with
the positive and
negative energy solutions.\,\footnote{The operators $a_0$ and $a_0^\dagger$ are not
necessarily required to be particle annihilation and creation operators, see
Ref.~\cite{jackiw rebbi} for details.} The first term in the expansion
(\ref{zero mode full expansion}) is the zero mode.
 Inserting the expansion (\ref{zero mode full expansion}) into Eq.
(\ref{charge definition}), we get
\begin{eqnarray}
\label{charge expansion}
    \langle q\rangle
    &=&
    -\frac{1}{2}\langle a_0 {a_0}^\dagger-{a_0}^\dagger  a_0\rangle
    \nonumber\\[3mm]
       &-&\!\!\!\!
       \sum_{\begin{array}{ll}
n\neq 0\\
\sigma=1,2
\end{array}} \left\langle
a_{n,\sigma}^\dagger a_{n,\sigma}-b_{n,\sigma}^\dagger b_{n,\sigma}
-a_{n,\sigma} a_{n,\sigma}^\dagger+b_{n,\sigma}b_{n,\sigma}^\dagger
\right\rangle
\end{eqnarray}
The condition we impose on $a$ is $a|{\rm vor}\rangle=0$. With this
condition we get
\begin{equation}\label{charge final answer}
\langle q\rangle=-{1}/{2},
\end{equation}
which again agrees with the previous results.
\\
\mbox{}
\\
\mbox{}

\end{document}